\documentclass[prd,aps,floats,tabularx,preprintnumbers,twocolumn]{revtex4}
\usepackage{graphicx, epsfig}

\begin{document}

\baselineskip 12 pt

\title{Does the Low CMB Quadrupole Provide a New Cosmic Coincidence
Problem ?}

\author{ Alessandro Melchiorri$^*$, Laura Mersini-Houghton$^\ast$}

\affiliation
{$^\ast$ Department of Physics and Astronomy, UNC-Chapel Hill, CB 3255 Phillips Hall, Chapel hill, NC 25599-3255, USA\\ 
$^*$ Physics Department and INFN, University of Rome
``La Sapienza'', Ple Aldo Moro 2, 00185, Rome, Italy\\}

\begin{abstract}
We point out that the recent WMAP findings of a suppressed large scale 
power in the Cosmic Microwave Background anisotropy power spectrum, if of cosmological 
origin, make the case for a new cosmic coincidence problem.The observed suppression of anisotropies at large angles requires an explanation for why this suppression is occuring {\it only} for perturbation wavelengths of the order 
of our present Hubble horizon. This is a very challenging task since a consistent model for the emerging picture of the universe has to address both cosmic coincidences and thus may reveal a deeper 
inter-relation with the dark energy problem.
\end{abstract}

\maketitle

Recent developments in precision cosmology have presented theoretical
physicists with a tantalizing picture of the universe. By a combination of
all data, there is indisputable evidence that our universe is
accelerating\cite{spergel}. This means that about $70\%$ of the energy
density in the universe is made up by a mysterious component, 
coined dark energy, with an equation of state $-1.38 \leq w_X \leq -0.82$ 
at $95\%$ confidence level (see e.g. \cite{melchiorri}).

Cosmic microwave background (CMB) measurements have provided valuable data that have proven a powerful tool
in lending support to a concordance $\Lambda CDM$ picture in cosmology. Nevertheless
we still lack an understanding of the origin and nature of DE and dark
matter which together make for about $95\%$ of the energy density in the
universe's budget.

It is well known that the dark energy (DE) mystery is a notoriously difficult problem, 
mainly due to the following two fundamental questions which need to be addressed simultaneously: 
why is DE magnitude $\rho_{X} \simeq 10^{-122} M_P^4$,  122 orders less than the
expected value $M_P^4$. This is known as the fine-tuning problem; why is DE domination 
over matter energy density in driving the expansion of the universe occurring around redshifts
 $z\simeq 0-1$ i.e at scales of the order the present value of the Hubble radius $H_0 \simeq
10^{-33} eV$. The latter is known as the coincidence problem of DE\cite{desean} and 
hereafter we refer to it as {\it the first cosmic coincidence}. 

In this work we discuss the possibility that the CMB data accumulated recently,  reveals a 
{\it second cosmic coincidence} stemming from the suppressed perturbation modes 
crossing the horizon only recently, and at the same energy scale as the DE coincidence.
The recent results from the WMAP satellite experiment on the Cosmic Microwave Background 
anisotropy power spectrum have been an extraordinary success for the standard model
of structure formation based on dark matter, inflation and dark energy (see e.g. 
\cite{wmap}). These results have not only confirmed (with a sensible reduction of the 
error bars!) the previous cosmological picture as defined by several ground-based 
and balloon-borne experiments (see e.g. \cite{netterfield}) but also presented 
indications for several intriguing discrepancies. 
Between the most surprising findings, WMAP measured a 
suppression of power at large angles, (low multipoles $l$), for
temperature auto-correlations $C_l^{TT}$ in the CMB anisotropy spectrum\cite{wmap}.
The indication for a quadrupole suppression was already reported a decade ago by the $COBE/DMR$ 
experiment but with much lower signal-to-noise and frequency coverage.
The higher frequency coverage of the WMAP experiment has, for example, excluded an
explanation of the effect as a systematic related to known galactic foregrounds.

Moreover, recent analysis \cite{costa} of the WMAP maps 
have reported that the cosmic quadrupole ($l=2$) on its 
own is anomalous at the $1$-in-$20$ level by being low, the cosmic octopole ($l=3$) 
on its own is anomalous at the $1$-in-$20$ level by being very planar and that the 
alignment between the quadrupole and octopole is anomalous at the $1$-in-$60$ level.
While the large scale temperature power spectrum clearly shows
discrepancies, other large scale CMB measurements are providing interesting clues.
For example, the measured level of large angle cross temperature-polarization 
$C_l^{TE}$ is higher than what expected in the most common scenario of 
reionized intergalactic medium (see e.g. \cite{kogut}). 
Furthermore, recent works \cite{nongaus} have presented evidence for large scale 
non gaussian-signals in the WMAP maps which again, if of cosmological origin,
would present a problem for the standard scenario.



\begin{figure}[thb]
\begin{center}
\includegraphics[scale=0.4]{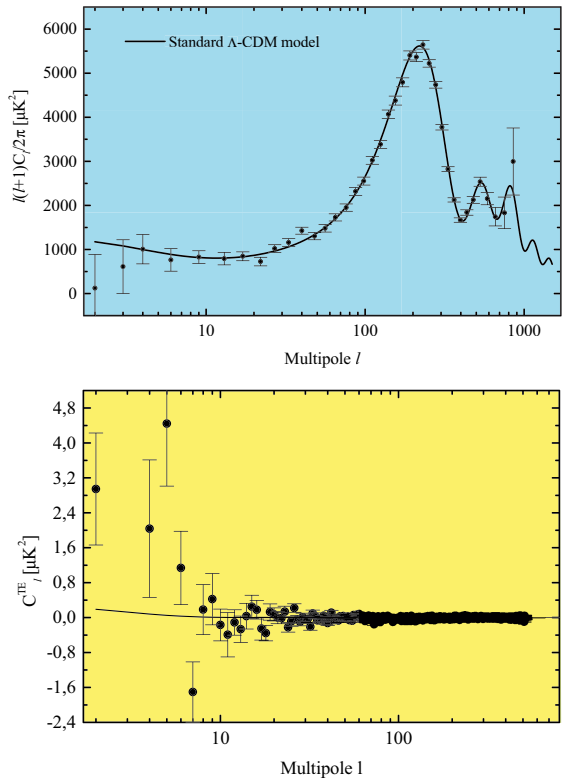}
\end{center}
\caption{The CMB power spectra from the 1st year WMAP data release for
temperature (top panel) and cross temperature-polarization
(bottom panel) compared with the expectations of a standard $\Lambda$-CDM 
model. The lack of power at large scales (small $\ell$)
in the temperature maps can be interpreted as a signature for new physics.
The high power at large scale in the $TE$ spectrum is in tension with 
the expected values (based on numerical simulations) of $\tau \sim 0.05-0.10$.It might suggest an higher 
value for the optical depth of the universe ($\tau \sim 0.2$)}
\label{mmog}
\end{figure}

These findings cannot be considered as conclusive evidence yet because of the
limitations set by cosmic variance or by possible systematics
(see e.g. \cite{efstathiou}). Statistical ``flukes'' are indeed present 
at several multipoles. Furthermore, the correlations with the local 
dipole and ecliptic may suggest presence of systematics due, for example, 
to unknown  foregrounds or present during the in-flight dipole calibration
(see e.g. \cite{schwarz}).

However they are intriguing enough to
motivate further effort in circumventing cosmic variance. This can be 
achieved by means of complimentary data like cosmic shear from weak 
lensing\cite{shear} and cross-correlations with the polarization 
spectra that are lending support to $WMAP$ findings, (see e.g. \cite{cross}, \cite{cosmicvar}).

Thus, taking the origin of the large scale CMB suppression as cosmological, 
presents a new cosmic coincidence because it requires that we address two questions with 
respect to the observed suppression of low 'l' CMB modes: 1) Why are these modes suppressed and, 
2) Why is the suppression 
occuring only for perturbation wavelengths of the order of our present Hubble horizon
$\lambda_0 \simeq H_0^{-1} \simeq 10^4 Mpc, k_0\simeq H_0\simeq 10^{-33}eV $.
 Analogous with the DE issues, let us refer to the first puzzle as the low 'l' tuning problem of 
CMB and to the second puzzle as the cosmic coincidence of CMB. 
(Notice that by tracing the CMB modes back in time the suppression of anisotropies at 
large angular scales may introduce a tuning in the number of inflationary efodings by 
imposing a cuttoff in their maximum number, \cite{linde}. 
This motivates the name 'tuning' for the first CMB puzzle). 

A lot of effort has recently been focused in trying to address the first puzzle namely, to 
explain the reason why the CMB low 'l' modes are suppressed. But the second question about the 
observed CMB suppression at low ' l' namely, the {\it 'why now'} puzzle,  which reveals a new 
cosmic coincidence in our present universe, has not been investigated until now. 
The fact that we have both cosmic coincidences, DE and CMB occuring at the same energy scale 
remains also to be addressed. Finding a mechanism that would suppress the large scale modes 
is not enough because, for a consistent picture of the present universe, we also have 
to address why the suppression occurs at the DE scale, $H_0 \simeq 10^{-33} eV$. 

It is reasonable to expect that both phenomena may originate from processes 
occurring in the very early universe, despite that these coincidences associated with the 
two currently observed phenomena namely, DE domination and CMB power suppression at 
present-horizon sized wavelengths, 
are dominantly displayed at low energies, for the following reasons:  
Recall that in an inflationary universe perturbations produced near the end of inflation
leave the horizon whenever their wavelength becomes larger than
the inflationary horizon $H_i$ due to 'superluminal' propagation. These
modes re-enter the horizon at later times when the Hubble parameter once
again becomes equal to their wavelength. This is known as the horizon crossing
condition $k = a(t) H(t)$. Thus the largest wavelengths are the first ones
to leave the horizon and the last ones to re-enter. Modes currently
re-entering $k_0 =a_0 H_0$ have wavelengths horizon size,which means they 
have been outside of the Hubble horizon for most of the history of the universe.
Thus they have not been contaminated by the internal evolution and
nonlinearities of the cosmic fluid inside the Hubble radius. These modes
carry the pristine information of the unknown physics which sets the Initial
Conditions of the universe\cite{katie}.

The possibility of a late-time effect that DE may have on the suppression of 
large angle anisotropies still remains open\cite{alesilklm} although it does not 
look like a promising direction at the moment . So far there has been no succesful 
attempt that can accomodate and relate both 
coincidences within the framework of low energy or late time physics. 
Perhaps this is because, based on the inflationary paradigm, we expect a vacuum energy 
component to enhance power of long wavelengths due to the integrated Sachs Wolfe effect (ISW).

What kind of physics could give rise to a consistent picture ? Two coincidences in our present universe can not be accidental.Perhaps this bizarre picture of the present universe emerging from precision cosmology is providing clues of new physics.
String theory and quantum gravity are possible candidates of the unknown
physics of the early universe. There are current models in literature that
offer an explanation for the CMB power suppression, by having the
Initial Conditions set within the framework
of string theory\cite{katie,dvali}, loop quantum
gravity\cite{marteens}, or an unknown hard cutof\cite{linde}. 
However none of these models can address the issue as to why the suppresion of anisotropies 
in the visible spectrum should occur at exactly {\it only} those wavelengths of order our present Hubble 
radius and not for any other modes that crossed the horizon earlier, 
(shorter wavelengths), although we have an infinite range of shorter wavelength modes.
 
There is also an ongoing search for a possible $UV/IR$
mixing of gravitational scales\cite{banks}. The motivation comes from the fact that both 
observed cosmic coincidences occurring at present low energy scales, namely the DE coincidence 
and CMB coincidence, independently suggest constraints imposed on the number of efoldings, 
(about $60 efoldings$ for GUT scale inflation), during the early times when the universe underwent 
an inflationary episode \cite{banks,katie}. However a theoretical model that can successfully 
accommodate all observed cosmic coincidences around the scale $H_0 \simeq 10^{-33}eV$ is yet 
to be found.

A single solution to the DE and low CMB quadrupole problems will have to address both 
cosmic coincidences and also reveal a deeper inter -relation between the two, a very challenging task. 

It is worth mentioning that recent works have presented evidence for correlations between the large-scale WMAP 
data and radio sources distributions \cite{nolta} which certainly provide hints that the suppressed 
power may be connected to the recent DE domination.

An important implication of having two cosmic coincidences at the same scale is the possibility that this energy scale is 
{\it special}, possibly indicating that our present Hubble 
parameter may turn out to be a {\it new scale of physics at very low energies}.
Or perhaps a new scale of low energies derived from a fundamental scale of high
energy physics through a possible $UV/IR$ mixing. 
This radical possibility, although very intriguing, is not yet realized in a concrete model.

At the moment, our theoretical knowledge of the relation between 
the low CMB quadrupole and the dark energy lies in the realm of speculations 
while pushing forward the discovery of new physics.
Future investigation of the {\it new cosmic coincidence} puzzle discussed here, may 
independently aid us to a better understanding of the emerging picture of the present universe.

Acknowledgement: LM is very grateful to the hospitality of  Perimeter Institute for 
Theoretical Physics, Waterloo, ON Canada N2J 2W9, during her visit there where part of this work 
was done and especially to many stimulating discussions with L.Smolin.

\end{document}